\documentstyle[11pt]{article}
\def\lsim{\mathrel{\rlap{\lower4pt\hbox{\hskip1pt$\sim$}}
    \raise1pt\hbox{$<$}}}                
\def\gsim{\mathrel{\rlap{\lower4pt\hbox{\hskip1pt$\sim$}}
    \raise1pt\hbox{$>$}}}                

\begin{document}
\def\beq{\begin{equation}}
\def\eeq#1{\label{#1}\end{equation}}
\def\barr#1{\begin{equation}\begin{array}{#1}\displaystyle}
\def\earr#1{\end{array}\label{#1}\end{equation}}
\def\bit{\begin{itemize}}
\def\eit{\end{itemize}}
\def\ben{\begin{enumerate}}
\def\een{\end{enumerate}}
\def\bce{\begin{center}}
\def\ece{\end{center}}
\def\bmi{\begin{minipage}}
\def\emi{\end{minipage}}
\def\btab{\begin{tabular}}
\def\etab{\end{tabular}}
\def\quo#1{\begin{quote}#1\end{quote}}
\def\dis{&\displaystyle}
\def\di#1{\\[0.#1cm]\displaystyle}
\def\av#1{$<\!\!#1\!\!>$} 
\def\avv#1{<\!\!#1\!\!>}
\def\r#1{(\ref{#1})}
\def\df{\stackrel{\rm def}{=}}
\def\ot#1#2{\textstyle \frac{#1}{#2}}
\def\od#1#2{\displaystyle \frac{#1}{#2}}
\def\lil#1{\hbox to\hsize{ #1 \hfil}}
\def\lilr#1#2{\hbox to\hsize{#1 \hfil #2 }}
\def\lir#1{\hbox to\hsize{\hfil #1 }}
\def\lic#1{\hbox to\hsize{\hfil #1 \hfil}}
\def\v#1{\vspace{#1cm}}
\def\h#1{\hspace{#1cm}}
\def\hh#1{\vrule height.4pt width#1cm depth-.2pt \ }
\def\hhh{\vrule height.4pt width\hsize depth-.2pt \ }
\def\si{\sigma}
\def\ga{\gamma}
\def\ep{\epsilon}
\def\al{\alpha}
\def\be{\beta}
\def\de{\delta}
\def\De{\Delta}
\def\La{\Lambda}
\def\la{\lambda} 
\def\ta{\tau}
\def\Z{\indent\indent}
\def\N{$\cal N$}

\newread\epsffilein    
\newif\ifepsffileok    
\newif\ifepsfbbfound   
\newif\ifepsfverbose   
\newif\ifepsfdraft     
\newdimen\epsfxsize    
\newdimen\epsfysize    
\newdimen\epsftsize    
\newdimen\epsfrsize    
\newdimen\epsftmp      
\newdimen\pspoints     
\pspoints=1bp          
\epsfxsize=0pt         
\epsfysize=0pt         
\def\epsfbox#1{\global\def\epsfllx{72}\global\def\epsflly{72}%
   \global\def\epsfurx{540}\global\def\epsfury{720}%
   \def\lbracket{[}\def\testit{#1}\ifx\testit\lbracket
   \let\next=\epsfgetlitbb\else\let\next=\epsfnormal\fi\next{#1}}%
\def\epsfgetlitbb#1#2 #3 #4 #5]#6{\epsfgrab #2 #3 #4 #5 .\\%
  \epsfsetgraph{#6}}%
\def\epsfnormal#1{\epsfgetbb{#1}\epsfsetgraph{#1}}%
\def\epsfgetbb#1{%
%
%
\openin\epsffilein=#1
\ifeof\epsffilein\errmessage{I couldn't open #1, will ignore it}\else
 {\epsffileoktrue \chardef\other=12
    \def\do##1{\catcode`##1=\other}\dospecials \catcode`\ =10
    \loop
       \read\epsffilein to \epsffileline
       \ifeof\epsffilein\epsffileokfalse\else
       \expandafter\epsfaux\epsffileline:. \\%
       \fi
   \ifepsffileok\repeat
   \ifepsfbbfound\else
   \ifepsfverbose\message{No bounding box comment in #1; using
defaults}\fi\fi
   }\closein\epsffilein\fi}%
\def\epsfclipon{\def\epsfclipstring{ clip}}%
\def\epsfclipoff{\def\epsfclipstring{\ifepsfdraft\space clip\fi}}%
\epsfclipoff
\def\epsfsetgraph#1{%
   \epsfrsize=\epsfury\pspoints
   \advance\epsfrsize by-\epsflly\pspoints
   \epsftsize=\epsfurx\pspoints
   \advance\epsftsize by-\epsfllx\pspoints
 \epsfxsize\epsfsize\epsftsize\epsfrsize
  \ifnum\epsfxsize=0 \ifnum\epsfysize=0
      \epsfxsize=\epsftsize \epsfysize=\epsfrsize
      \epsfrsize=0pt
 \else\epsftmp=\epsftsize \divide\epsftmp\epsfrsize
       \epsfxsize=\epsfysize \multiply\epsfxsize\epsftmp
      \multiply\epsftmp\epsfrsize \advance\epsftsize-\epsftmp
       \epsftmp=\epsfysize
       \loop \advance\epsftsize\epsftsize \divide\epsftmp 2
       \ifnum\epsftmp>0
          \ifnum\epsftsize<\epsfrsize\else
             \advance\epsftsize-\epsfrsize \advance\epsfxsize\epsftmp \fi   
       \repeat
       \epsfrsize=0pt
     \fi
   \else \ifnum\epsfysize=0
     \epsftmp=\epsfrsize \divide\epsftmp\epsftsize
     \epsfysize=\epsfxsize \multiply\epsfysize\epsftmp
     \multiply\epsftmp\epsftsize \advance\epsfrsize-\epsftmp
     \epsftmp=\epsfxsize
     \loop \advance\epsfrsize\epsfrsize \divide\epsftmp 2
     \ifnum\epsftmp>0
        \ifnum\epsfrsize<\epsftsize\else
           \advance\epsfrsize-\epsftsize \advance\epsfysize\epsftmp \fi
     \repeat
     \epsfrsize=0pt
    \else
     \epsfrsize=\epsfysize
    \fi
   \fi
   \ifepsfverbose\message{#1: width=\the\epsfxsize,
height=\the\epsfysize}\fi
   \epsftmp=10\epsfxsize \divide\epsftmp\pspoints
   \vbox to\epsfysize{\vfil\hbox to\epsfxsize{%
      \ifnum\epsfrsize=0\relax
        \includegraphics{\ifepsfdraft}%
      \else
        \epsfrsize=10\epsfysize \divide\epsfrsize\pspoints
        \includegraphics{\ifepsfdraft}%
      \fi
      \hfil}}%
\global\epsfxsize=0pt\global\epsfysize=0pt}%
{\catcode`\%=12
\global\let\epsfpercent=
\long\def\epsfaux#1#2:#3\\{\ifx#1\epsfpercent
   \def\testit{#2}\ifx\testit\epsfbblit
      \epsfgrab #3 . . . \\%
      \epsffileokfalse
      \global\epsfbbfoundtrue
   \fi\else\ifx#1\par\else\epsffileokfalse\fi\fi}%
\def\epsfempty{}%
\def\epsfgrab #1 #2 #3 #4 #5\\{%
\global\def\epsfllx{#1}\ifx\epsfllx\epsfempty
      \epsfgrab #2 #3 #4 #5 .\\\else
   \global\def\epsflly{#2}%
   \global\def\epsfurx{#3}\global\def\epsfury{#4}\fi}%
\def\epsfsize#1#2{\epsfxsize}
\let\epsffile=\epsfbox


\vspace{2cm}

\lic{\Large \bf Concept of Powerful Multistage Coaxial}
\vspace{.2cm}
\lic{\Large \bf Cyclotrons for Pulsed
and Continuous Beam Production. } 
\vspace{.5cm}
\vspace{.5cm}
\lic{A.R.~Tumanyan,  N.Z. Akopov,  Z.G. Guiragossian,  Z.N. Akopov}
\vspace{0.5cm}
\begin{center}
Yerevan Physics Institute, Yerevan, Armenia
\end{center}
\vspace{2cm}

\begin {abstract}
The concept of large-radius multistage coaxial cyclotrons having
separated orbits is described, to generate proton beams of 120-2000
MeV energy at tens of GW pulsed and hundreds of MW in continuous beam
power operation.  Accelerated beam losses must be less than 0.1 W/m for
the intercepted average beam power linear density.  

The concept is
inherently configured to actively compensate the longitudinal and
transverse space charge expansion in beam bunches. 
These are
based on (1) actively varying the bunch acceleration equilibrium phase
while maintaining isochronism, independently for each cyclotron turn;
(2) independently changing the acceleration voltage for each turn  
together with orbit corrections that preserve isochronism; (3)
independently changing the transverse betatron oscillation tune
shift, to assure
non-resonant operation.  Also, (4) sextupole lenses are included 
to compensate for chromaticity effects.
Moreover, the concept is based on optimum uses of practical successful  
results so far achieved in beam acceleration and storage techniques. 

These are the reasons why the proposed multistage cyclotron system   
appears doable.  This accelerator has a wide range of applications.
It can especially be used to deliver a pulsed intense source of neutrons
for scientific research without the use of follow-on storage rings for
pulse compression, and also, to drive the industrial transmutation
technologies. 

As an example of such a cyclotron system, we describe our approach of   
accelerating single-bunch proton beams at up to 1 GeV energy, with    
pulsed beam power of 80 GW and bunch duration of 2 ns. The exemplar
cyclotron accelerator system is configured to be located in the shielded
structure of the 6-GeV Yerevan Electron Synchrotron. The cost of such a
cyclotron system is estimated to be approximately 40,000,000 US dollars,
if implemented in Armenia at substantially reduced labor costs.

\end {abstract}


\section{Introduction} \Z
High power proton or ion accelerators at hundreds MW in    
continuous mode are needed for the Accelerator Driven Transmutation
Technologies ({\bf ADDT}), in different nuclear power production
industrial   
applications.  However, the same cyclotron accelerator system can also 
be applied to produce ultra-short beam pulses for scientific
investigations, for the production of powerful neutron sources.
In this case, generated pulsed beams may have peak power in the range
of 10-100 GW.  If in addition the duration of ejected pulsed beams is
less than 1.0 ms at energies of 1.0 GeV, it is also possible to exclude
the necessity of placing a last-stage storage ring for pulse compression
to produce intense neutron bursts [1,2].

This work is performed to optimize plans for the construction of a      
proton accelerator system with 1.0 GeV energy and tens of mA current,   
to be realized in the existing shielded structure of the Yerevan 6-GeV
Electron Synchrotron. The primary approaches, based on the use of
large-radius isochronous Coaxial Ring Cyclotrons (CRC) and Asynchronous
Cyclotrons (AC) were described in references [3-7].

The key problems that relate to the construction of powerful proton
accelerators are the following:
- To greatly decrease the beam losses
- To increase the efficiency and reliability of accelerator's with
reduced operating costs
- To decrease the development and construction costs.

The most difficult problem is to get high values of average beam current
with acceptably low values of accelerated beam losses.

\section{Concept of multistage coaxial cyclotrons} \Z

The concept of Multistage Coaxial Cyclotron {\bf (MCC)} gives an opportunity
to perform flexible management of separated cyclotron orbits in the     
energy range of 120-2000 MeV, and to also obtain hundreds of mA average
beam current in continuous beam operation.  These advantages are based
on demonstrated accelerator physics and technologies that have been
implemented predominantly on powerful synchrotrons and linear
accelerators (linacs).

We first note that on HERA [8,9] it has been demonstrated that injection
and acceleration of proton beams in synchrotrons (without storage)
proceed well when the number of particles in each bunch is limited to
$(1.2 \div 1.4) \times 10^{11}$, for acceleration frequencies of 10-100 MHz and energies 
of tens of MeV to tens of GeV.

Second, in the same publications it is shown that due to high-precision
strong focusing and high vacuum, beam losses during many years of HERA
machine operation have not exceeded the tolerable value of 0.01 nA/m, 
for energies higher than 100 MeV.

Third, in modern high power linacs [10,11], the construction, 
transmission and handling of megawatt level RF power from single sources
have been demonstrated at frequencies of 350-700 MHz.  In principle, it 
is also possible to develop similar approaches for large acceleration
cavities [12,18] and low frequencies of 50-100 MHz.

It also appears that in a large radius {\bf MCC} it would not only be possible
to create conditions like those existing in proton synchrotrons for  
strong longitudinal and transverse beam focusing, but to attain much
better conditions due to the possibility of the more flexible and
relative independent management of the separate orbit parameters.

Thus, if we set the acceleration field of cyclotrons to have a frequency
in range of 50-100 MHz and as obtained at HERA, limit the number of  
particles per bunch to $(1.2 \div 1.4) \times 10^{11}$, in continuous mode of operation
an average acceleration current of up to one Ampere
$[(1.2 \div 1.4) \times 10^{11} \times (50 \div 100) \times 10^{6}]$ should be feasible for 
reasonably low beam interception losses.

In order to realize such a {\bf MCC}, the structure, the number of stages and 
their      
parameters will be  chosen according to  the following considerations:

1. In order to provide for flexible longitudinal beam 
focusing, the
opportunity of independently tuning in a wide range of up to tens of
degrees
of
equilibrium phase in acceleration fields will be exercised, in one or two
pairs of 
turn sectors, in each separate orbit turn while maintaining isochronism.

2. To create a strong and quality transversal focusing of beam, the
providing
of enough space between the accelerating cavities to install the warm
magnet
system with separated functions and with additional sextupole lens to
correct
the chromaticity and also to change the mechanical coordinates of magnet 
elements in the range of   
several millimeters and length of bending magnets in the range of several
centimeters.

3. In order to increase the efficiency and simplify the injection and
ejection of beams, the parameters of {\bf MCC} stages is tuned to be possible to
put these stages coaxial or concentric, each into the other.

4. In order to increase the opportunities of beam parameter's tuning, the
value of the turn separation is chosen to be more than 20cm. and
camera's aperture, not less than $ 3 \times 5cm $.

5. To compensate the shift of bethatron oscillations frequency 
from the effect of the beam space charge, the opportunities of slow
changing
of Q value from one magnet period to the other in according
with computer
code
and depending of measured beam parameters, are supposed.

6. The cavities could be warm as well as superconducting ({\bf SC}). The
choice
depends on a lot of reasons and the firstly on opportunities and
tastes of desingners.

7. In order to increase the opportunities to compensate the mutual
influence
of beams into the accelerating cavities and spread of cavities' parameters
during the accelerator tuning, besides of conditions 1, the opportunities
of
independent changing of
accelerating voltage on each beam's channel, using not mechanical methods
is  
supposed.

8. It's reasonable to create the {\bf MCC} on energies more than 120MeV.
Lower
energy accelerators seem to be not satisfied from technical point of view,
because of essential decreasing the "RF acceptance" value and increasing
of
the tolerances on the magnetic field's quality. That becomes more
remarkable
at high values of harmonic h of accelerating RF field, however, on the
other
hand the stable operation of accelerator usually is easily realized on
super 
narrow phase width of accelerating beam. The upper value of energy does
not
limited, because of formal technical reasons are absent and only
economical
reasons could be mainly considered, acceptable limit for which seems to be    
around 2000MeV.

9. Nowadays opportunities to construct the proton accelerators with
energy up
to 120MeV are wide enough, starting the RFQ and finishing the {\bf SC}
Cyclotrons
with Separated Orbits. That's why the decision of question
concerning
the Project of injecting system with final energy up to 120MeV depends on
opportunities and tastes of designers and will be considered separately.

10. The price of construction and exploring of {\bf MCC} should be essential
lower 
the price of adequate Linac.

11. Developing of {\bf MCC} Project must be also based on enough studying of
experience and results of designing and construction of similar cyclotrons    
with separated orbits \cite{russel,brovko}.

\section{MCC pulsed operation mode} \Z
        Anyway, even constructing the {\bf MCC} with continuous functioning,
it's
reasonable to have opportunity to realize initial adjustment and start up
of
accelerator in pulsed mode and rather with acceleration of one-bunch beam, 
which is formatted into the injector's structure. The radius of the last
stage
of {\bf MCC} is chosen from the condition to provide the duration of one turn of
beam to be less than 1.0 mksec., in order to satisfy the requirements
applied
to using proton beam into the system of neutron source. Thus, if the peak     
value of current for accelerating beam in {\bf MCC} will be high enough (about
tens
Ampere), then it's not necessary to create the storage rings \cite{Jose,
Austron}.

        During one-bunch regime of {\bf MCC} operation, in any time only one
bunch
will pass any cavity, and time interval between loading the cavity by beam 
(bunch) is equal the duration of one turn of beam at given stage.
Instabilities at such regime of operation and essential limitation of
intensity (took place in regime of continuous acceleration of beam), 
connected
with these instabilities , are just disappeared.

        So, during the short pulsed feeding of cavity (20-40mksec.), 
depends on
the level of connection with the RF generator and quality factor
, it's possible to
increase
significantly the value of electrical field in cavity(before appearance of
multiplicity effects) and use the warm cavities.

        However, in the same time it seems to be not technically
profitable
the construction of pulsed feeding for Magnet Elements ({\bf ME}- bending
magnets
and lenses) during the period of the pulsed mode accelerating in {\bf
MCC}.
The
explanation is, that for pulsed feeding of {\bf ME} the distortions and
instabilities of the {\bf ME's} magnetic fields are dramatically
increased. This
is 
connected mainly with drifting of "coercivity"  of the magnet's
core iron, and demagnetization by inverse polarity current is not  
enough stabilizing effect, particularly for the edge of magnet and short
lens.

Besides that, the pulsed feeding determines the necessity of {\bf ME}   
construction, using more thinner of sheet steel, that leads to the
increasing
of mechanical distortions of the core. Moreover, the isolation of
winding
increases, that leads to the according increasing of the {\bf ME} sizes.
As a
result,
 the opportunities to construct small sizes {\bf ME}, which are very
important for
considered type of cyclotron, are decreased. So, the {\bf ME} feeding
system
becomes
complicated, particularly the possibilities of precise stabilization of
this  
system. The main thing is, that identity of such magnet system will be
destroy
during the transition from pulsed to continuous feeding.

That's why the chose of continuous current for the {\bf ME} feeding,
during the
pulsed one-bunch beam acceleration in {\bf MCC}, will allow to move smoothly to 
multi-bunches acceleration up to continuous regime. That is possible only
at 
the expense of the RF system power increasing and seems to be technically  
justified.

        Anyway, initial tuning of the {\bf MCC} with one-bunch acceleration in
each
cycle of accelerating open also some additional opportunities to increase
the 
quality of cyclotron. For example, it's possible to identify by help of
the   
same bunch all detectors and also some
others   
parameters of cyclotron in all channels and stages. Besides that, 
cycling rate  can be not only 50-60Hz, but will be varied in a
wide
range
from zero up to several thousands Hz.

        Described advantages of one-bunch acceleration 
, will allow, in roughly estimation, to increase the  
number of particles in single bunch 
more than in one order, e.g. to reach the values of $10^{12}$ and more.
Taking in
account, that phase duration of bunch is usual about 
0.1 of
oscillations period of accelerating RF field in cavity, then, at the RF
oscillations frequency about 50MHz, the pulsed power of ejected  {\bf MCC}
beam 
at energy about 1GeV will be about 80 GW with pulse duration equal to 2ns.
It 
seems to be enough to construct the powerful pulsed neutron source
without
the storage ring \cite{Jose,Austron}.

        For the future, 
with each increasing of bunches number, the
number  
of stable accelerating particles in each bunch will be a little decreased.    
This process will be continued till the reaching of the continuous mode,
after that
process 
will mainly finish. However, if {\bf MCC} system is constructed and adjusted
correctly, then the number of particles located in each bunch at
continuous  
regime should not be less the value already reached on practice, e.g.
$10^{11}$.

        With increasing of average current of beam it's necessary to have   
mainly according increasing of accelerating RF system power, what is
mainly
connected with financial opportunities. Thus, one more practical advantage
of 
suggested type of accelerator is the possibility of the gradual upgrade
of
it
power without changing of structure and magnet system, any upgrade
depends  
just on financial opportunities.

        It will be not so correct to estimate generally the cost of {\bf
MCC}, 
because of strong dependence on the "making" country. Because of that, we
will
initially consider some relative cost reasons in comparison with Linac, 
which
seemed to be independent enough from the "making" country.

        For example, the length of Linac at 1GeV energy in SNS and
{\bf ADDT}     
systems \cite{Jose,Austron,jason,lawrence} is about 1.0 km. Then
the length of shielding concrete
wall
of 3m thickness will be about 2km. While, the same external wall for {\bf MCC}
with 
similar parameters of beam and with final radius about 38m(see bellow), 
will
have a length almost in one order shortly. One can expect, that also the
cost 
of expensive wall will decrease about ten times. Besides that, the cost of
accelerating RF technique on 50MHz frequency is usually significantly
lower
than for similar technique on 350 and 700MHz frequencies.
        Approximate picture of cost values for suggested type accelerator
construction one can image, based on the estimations of four-stage {\bf MCC}
construction at 1GeV energy for one-bunch accelerating of beam at Yerevan   
Physics Institute (YerPhI) in to existing radiation zones of electron
synchrotron at 6GeV(see bellow). Based on results of these estimations we
can 
conclude, that construction of {\bf MCC} will take about 40,000,000 US dollars,
what
is more than one order less than cost of construction of similar
accelerators,
described in \cite{Jose,Austron,jason,lawrence}.
        We should note, that obtained estimation of cost does not include
the
cost of shielded structures, manpower and cost of 120MeV injection system
construction. The cost for injection system is estimated in about $ 15 \%$
of total
sum.
The manpower and shielded structures cost essentially depends on current
economical status in the "making" country.

\section{MCC based on YerPhI synchrotron}

The aim of this
work is to try to demonstrate the principal possibility and advantages of
a   
wide-applicable {\bf MCC} based on example, developed for YerPhi , relative to
similar accelerators.

The following data are used as a basis of calculations:
the main ring average radius is 34.5 m, the tunnel width is 6.0 m, with
3.0 m thick tunnel walls, and the inner circular hall diameter is 57.0 m.
As shown in Fig. 1, the first three cyclotron stages are nested and
concentrically
installed in the hall, acting as  injectors for the fourth
stage cyclotron in the {\bf YerPhI} main-ring tunnel.

        The choice of {\bf MCC} structure and its parameters depends on the
choice
of the type and parameters of the accelerating RF cavities. In
\cite{trinks,first,brovko}
the
research and design of large, low-frequency (90-170MHz) accelerating
{\bf SC}
cavities with no spherical symmetry and with substantial radial range of
gap are
described. In these cavities, as well as in large warm ones, which have
the
radial range of working gap more than 4m and 50MHz frequency
\cite{santafe,
willike}, the
inner active surfaces are made of Pb or PbSn(4Sn 96Pb atoms). At the
 same time, in warm cavities, the electrical fields of 6.2Mv/m are
reached,
and in {\bf SC} - 10.6Mv/m, while the ratio of the total electric field to
the
using
part of field is $E_{peak}/E_{acc}=1.5$.
        For the proposed four-stage {\bf MCC-YerPhI} the warm cavities' design
is 
taken into account, similar to those modeled in \cite{fietier} based on the
presently
operating in PSI \cite{santafe} modern cavities, at 40-50MHz frequencies and
achieving
peak voltage of up to 1.1MV, 
with radial range of working  gap up to 4m.
        The results of calculations for one of the versions of
{\bf MCC-YerPhI}, consisting of 4 rings are given in Table 1.  

\begin{figure}[h]
\begin{center}
      \mbox{
           \epsfysize=9cm
           \epsfbox{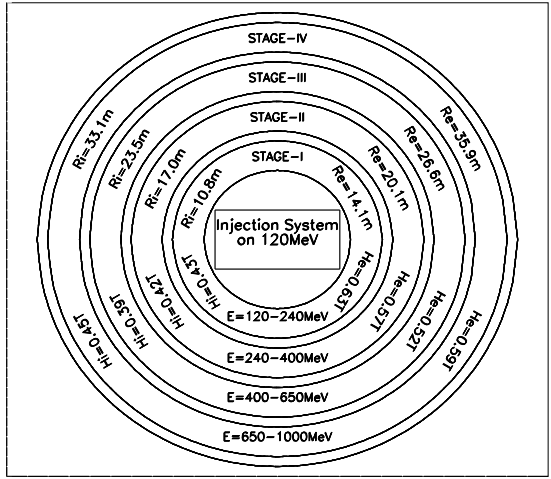}
           }
\end{center}
\caption{\it General layout of the Yerevan 1GeV {\bf MCC}
proton accelerator.}
\label{fig1}
\end{figure}

The transit-time factor value is assumed to 0.95.  
The magnet period structure is assumed as FODO lattice with
separate function.
\newpage
\begin{table}[h!]  
\caption{\it Four stages MCC-YerPhI parameters }
 \vskip 0.2cm
\begin{center}
\begin{tabular}{|l|c|c|c|c|} \hline \hline
PARAMETERS&STAGE-1&STAGE-2&STAGE-3&STAGE-4\\
\hline\hline
$E_{i}$-Injected Beam Energy [MeV]&120&240&400&650\\

$E_{e}$-Extracted Beam Energy [MeV]&240&400&650&1,000\\

$R_{i}$-Injection Beam Radius [m]&10.8&17.0&23.5&33.1\\

$R_{e}$-Extraction Beam Radius [m]&14.1&20.1&26.6&35.9\\

$N_{c}$-Number of Acceleration Cavities&10&15&22&40\\

$N_{m}$-Number of Sector Magnets&110&150&230&320\\

$N_{q}$-Number of Quadrupole Lenses&220&300&460&640\\

$N_{s}$-Number of Sextupole Lenses&220&300&460&320\\

H-Field in Sector Magnets [T]&0.43-0.63&0.42-0.57&0.39-0.52&0.45-0.59\\
G-Gradient in Quadrupoles[T/m]&1.65-1.82&2.47-2.78&2.98-3.56&4.20-5.30\\

$\Delta{E}$-Energy Gain per Turn [MeV]&11.0&16.5&24.2&44.0\\
$\Delta{R}$-Orbit Separation [cm]&39.9-23.5&40.0-24.9&39.7-22.4&46.9-26.1\\

Q-values&3.75&5.75&7.25&10.25\\
n-Number of Turns &11&10&10.5&8\\
h-Harmonic Number&25&30&35&43\\

$L_{m}$-Length of Sector Magnet [m]&2.3&2.3&2.3&1.5\\
$L_{f}$-Length of Drift Spase [m]&4.1-6.1&4.3-5.6&3.9-4.8&3.2-3.7\\
Total Weight of Sector Magnets [tonne]&178&243&372&328\\
\hline\hline
\end{tabular}
\end{center}     
\label{tab1}
\end{table}
\vskip -0.5cm

Table 1 shows, that the orbit separation  is in
the range of 40-23cm., and the length of drift space
is not less than 3.7m, which will allow 
not
only to place freely focusing magnet elements and detectors
, but
also to easily provide $100 \%$ beam ejection. The total value 
of
turns
in
the {\bf MCC-YerPhI} will be 39.5 .

Table 2 shows the computer calculations for  the
first turn of beam in the first stage of {\bf MCC}, at artificial change 
of equilibrium phase on +3.6 degree in sector 5 and on
-3.6 
degree in sector 6, while maintaining isochronism.
\begin{table}[h!]
\caption{\it Change of Bunch Equilibrium Phase versus Sector Magnetic Field Setting in the
First Turn of a Separate Orbit Cyclotron(SOC).}
 \vskip 0.2cm
\begin{center}
\begin{tabular}{|c|c|c|c|c|c|c|c|c|c|}
\hline
 &q=&dE&r[m]&R[m]&$L_{a}$[m]&$L_{b}$[m]&$L_{m}$[m]&[Tesla]&E\\
Sector\#&$h/N_{c}$&MeV/&Bending&Orbit&Left&Right&Sector&Field in&[MeV]\\
Around&&Cavity&radius&radius&Drift&Drift&Magnet&Sector&Beam\\
$1^{st}$turn&&&&&length&length&length&magnets&energy\\
\hline\hline 
1&2.5&1.1&3.73&10.852&2.367&2.247&2.344&0.439&121.09\\
2&2.5&1.1&3.73&10.894&2.385&2.256&2.344&0.441&122.19\\
3&2.5&1.1&3.73&10.932&2.392&2.274&2.344&0.443&123.29\\
4&2.5&1.1&3.73&10.973&2.410&2.283&2.344&0.445&124.39\\
5&2.51&1.1&2.41&11.011&2.844&2.729&1.516&0.693&125.49\\
6&2.49&1.1&5.12&11.057&1.992&1.850&3.217&0.327&126.59\\
7&2.5&1.1&3.71&11.083&2.429&2.347&2.337&0.453&127.69\\
8&2.5&1.1&3.73&11.135&2.477&2.317&2.344&0.454&128.79\\
9&2.5&1.1&3.73&11.160&2.448&2.371&2.344&0.456&129.89\\
10&2.5&1.1&3.73&11.211&2.501&2.343&2.344&0.458&131.00\\
\hline\hline
\end{tabular}
\end{center}  
\label{tab2}
\end{table}
\vskip -0.5cm

According to the evaluated errors in the dipole and quadruple alignment,
the
beam envelope, including orbit distortions, was computed. It sizes in
different stages of {\bf MCC} lies within the limits of $40mm \times 60mm$
and
determines
the aperture and outside dimension of magnets, the latter of which are
$20cm \times 20cm$. For the calculations the followings  values of
injecting
beam
emitance
were taken:
$E_{x}=15\pi \times mm \times mrad$ and $E_{z}=10\pi \times mm \times mrad
$.

Table 3 shows the main parameters of extracted beam and cost estimations
of
{\bf MCC-YerPhI} construction in two cases of accelerator functioning. The
continuous regime of hundreds MW beam production is not
considered,
because of low probability to  realize it at YerPhI, due to the financial
reasons. Still, one can easely obtain from these tables the main
information to
approximate more powerful continuous regime of accelerators' functioning.
\newpage
\begin{table}[h!]
\caption{\it Bunch Parameters of Separate Orbit Cyclotron in Single Bunch and Multiple Bunch
Operations and Related Cost Estimates }
 \vskip 0.2cm
\begin{center}
\begin{tabular}{|l|c|c|} \hline \hline
PARAMETERS&SINGLE BUNCH&MULTIPLE BUNCH\\
          &MODE&MODE\\
\hline \hline
Beam Specie&Proton&Proton\\
Beam Output Kinetic Energy&1.0GeV&1.0GeV\\
Number of Bunches per acceleration Cycle&1&25\\
Bunch Duration&2ns&2ns\\
Beam Pulse Length&2ns&500ns\\
Number of Protons per Bunch&$10^{12}$&$10^{12}$\\
Cycle Repetition Rate&50Hz&50Hz\\
Average Beam Power&8kW&200kW\\
Instantaneous Beam Current&80A&8A\\
Instantaneous Beam Power&80GW&8GW\\
Average Beam Loss Power &0.1W/m&0.1W/m\\
Total Electric Power&2.6MWe&2.7MWe\\
\hline \hline
COST ESTIMATION&Million US\$&Million US\$\\
\hline \hline
87 Acceleration Cavities, each at \$150k&13.05&13.05\\
RF Generators, at \$2.5M per MW(rf)&For 0.096 MW(rf)=0.24&For 0.29 MW(rf)=0.73\\
820 Bending Magnets, each at \$10k&8.20&8.20\\
2,920 Lenses, each at \$3k&8.76&8.76\\
Vacuum System&3.50&3.50\\
Accelerator Control System&2.50&2.50\\
Other Miscellaneous Items&1.50&1.50\\
\hline \hline
COST TOTAL&37.57&38.74\\
\hline \hline
\end{tabular}
\end{center}
\label{tab3}
\end{table}
\vskip -0.5cm
One can see for this type of machine, that it's profitable to choose the
low
values of RF accelerating frequency, because that makes accordingly easy
the
problem of precisions, almost for all systems, but on the other hand, the
problems to get the high values of electric fields into the cavities and
their
quality factor(Q) become complicated, although in {\bf SC} case this
effect is
weaker
displayed. 

Besides that,  high values of the harmonic number lead to the   
increasing of opportunities to tune of equilibrium phase and stability of   
acceleration. Therefore, the necessity to search the compromise decisions
is
obvious.
To increase the quality and stability of {\bf MCC} acceleration, the
installation of
sextupole lenses near the quadrupole lenses in the area of non zero
dispersion
orbit
to compensate the chromatity is foreseen. It's also foreseen to construct
the
electronic systems of inverse connection with controlled passing strip and
with feed of correcting voltage on special electrodes as well as
application
of others known methods. All these methods, including a new developed ones,   
will provide the minimal beam losses during the accelerating over the      
spiral orbit with varying orbit separation value.

\section{Conclusions} \Z
The main goal of given work is to demonstrate, that it's very important to
continue design, developing and improvement of such a new, perspective
type of
accelerator as {\bf MCC} complex, which will be useful not only for scientific
investigations, but also for nuclear industry.
In conclusion, the authors would like to thank the colleagues from YerPhI,    
particularly A.Ts. Amatuni    and  colleagues  from JINR (Dubna) for
useful  
discussions and remarks.

\end{document}